\documentclass[aps,pre,twocolumn,amsmath,amssymb,showpacs,amsfonts]{revtex4}
\usepackage{epsfig}
\usepackage{graphicx}
\usepackage{dcolumn}
\usepackage{bm}

\begin{document}

\title{Radiation entropy bound from the second law of thermodynamics}

\author{Itzhak Fouxon}

\affiliation{Raymond and Beverly Sackler School of Physics and Astronomy,
Tel-Aviv University, Tel-Aviv 69978, Israel}

\date{\today }

\begin{abstract}

It has been suggested heuristically by Unruh and Wald, and independently by Page, that among
systems with given energy and volume, thermal radiation has the largest entropy. The suggestion
leads to the corresponding universal bound on entropy of physical systems. Using a gedanken experiment
we show that the bound follows from the second law of thermodynamics if the CPT symmetry is assumed
and a certain general condition on matter holds.
The experiment suggests that a wide class of Lorentz invariant local quantum field theories
obeys a bound on the density of states.


\end{abstract}

\pacs{05.70.-a,65.40.gd,11.30.Er,05.70.Ln}

\maketitle

The demand that total entropy of a closed system never decreases imposes important constraints
on the structure of microscopic theories. A particularly interesting consequence of the demand is
the existence of universal upper bounds on entropy of physical systems. Such bounds can
be inferred by considering gedanken experiments that involve a reference thermodynamical system
interacting universally with systems from the considered class. The first example of such a
reference system was provided by black holes. Using that any physical system interacts with a black
hole, Bekenstein proposed a universal entropy bound
\begin{eqnarray}&&
S\leq \frac{2\pi E R}{\hbar c}, \label{a1}
\end{eqnarray}
where $S$ is the system entropy, $E$ is the energy and $R$ is the system size \cite{B3}. The bound is
based upon the demand that the total entropy never decreases in gedanken experiments that involve
system absorption by a black hole. The derivation
employs an extended version of the second law (the so-called generalized second law) that includes
the contribution of the black holes in the total entropy. Despite some objections on the realizability of
the considered gedanken experiment \cite{UW}, no physical counterexamples to the bound were found
since its proposition. Moreover, the bound was rederived by an alternative approach \cite{RB2}. Here
we wish to address another bound on entropy, related to a universal agent different from a black hole, namely
the radiation. In their discussion of the Bekenstein bound, Unruh and Wald \cite{UW,UW1} (see also \cite{PW}) and Page \cite{Page} proposed heuristically that at given energy and system size, the entropy of radiation is the largest so that
\begin{eqnarray}&&
S\leq S_{rad}(E, R)\propto\left(\frac{2\pi E R}{\hbar c}\right)^{3/4}, \label{a2}
\end{eqnarray}
with coefficient of proportionality of order one. Following \cite{B4} we will refer to the above
bound as the Page-Unruh-Wald (PUW) bound. The PUW bound expresses the expectation that at given $E$ and $R$ the state with the largest entropy is a gas (cf. \cite{Hooft}), and it takes into account that thermal radiation has the maximal concentration of about one particle
per volume with size of the Compton wavelength.  The bound is stronger than the Bekenstein bound and often the difference is large: the system size can be much larger than its Compton wavelength, $R\gg R_{Comp}\equiv \hbar c/E$. On the other hand, the bound (\ref{a2}) may fail in situations with strong gravity. For example, it fails for black holes that saturate (\ref{a1}).
Still, the bound can be expected to hold quite generally and for its use it is important to understand
what is needed for the bound to hold. Here we show that the second law of thermodynamics allows to shed light
on the underlying assumptions.

We consider a weakly self-gravitating thermodynamic system with given energy $E$, volume $V$ and entropy $S$.
We assume that the CPT symmetry holds and the antimatter system obtained upon the CPT transformation of the original system is also physically admissible. Then, the antimatter partner is in the same thermodynamic state as the original system and it has the same energy $E$, volume $V$ and entropy $S$. We now let the two systems, matter
and antimatter one, interact.
It can be assumed that before the moment of the interaction the two systems
are enclosed in a somewhat larger box, so that there is a spacing both between both the systems and the box.
\begin{figure}[ht]
\vspace{0.5cm}
\begin{tabular}{cc}
 \epsfxsize=6.0cm  \epsffile{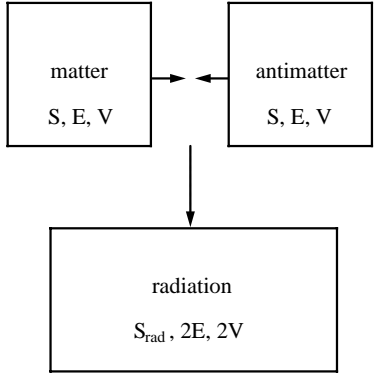}\\
\end{tabular}
\vspace{0.5cm}
\caption{The demand that the entropy does not decrease in the annihilation process
gives the radiation bound: initial entropy $2S$ of matter-antimatter couple must be not greater
than the final entropy of radiation at energy $2E$ and volume $2V$.} \label{eps0.8}
\end{figure}
The volume of the box can be chosen close to $2V$. Before the interaction the entropy inside the box is
given by (the entropy of the box is not essential in the following considerations):
\begin{eqnarray}&&
S_{in}=2S.
\end{eqnarray}
After the interaction the system and the "antisystem" annihilate giving rise to a certain product of the reaction.
Designating the entropy of the product by $S_{an}(2E, 2V)$, we find that the second law demand that the
entropy of the whole system does not decrease as a result of the reaction produces the inequality
\begin{eqnarray}&&
2S\leq S_{an}(2E, 2V). \label{boundgeneral}
\end{eqnarray}
In a wide class of situations the product of the annihilation reaction is the thermal radiation and
we will have
\begin{eqnarray}&&
S_{an}(2E, 2V)=S_{rad}(2E, 2V)=2S_{rad}(E, V),
\end{eqnarray}
where we used that $S_{rad}(E, V)\propto E^{3/4}V^{1/4}$, cf. Eq.~(\ref{a2}). The scheme of the experiment
in this case is shown in the Figure. Then the second law condition described by Eq.~(\ref{boundgeneral})
produces the sought for Page-Unruh-Wald bound
\begin{eqnarray}&&
S\leq S_{rad}(E, V)=g\left(\frac{E}{\hbar c}\right)^{3/4}V^{1/4}, \label{bound}
\end{eqnarray}
where the effective number of degrees of freedom (see \cite{Page,Hooft})
and numerical constants are absorbed in the $g-$factor which is of the order of a few.
It should be noted
that we use the term "thermal radiation" in the spirit of the original works \cite{UW,UW1}, where "radiation"
includes matter with the rest energy smaller or comparable with the temperature, see also \cite{Hooft}. In
particular, the $g-$factor in Eq.~(\ref{bound}) is energy density dependent near temperatures corresponding to a pair creation threshold.

We see that once the idea of the gedanken experiment is conceived, the demonstration of the bound is
rather immediate. A possible objection to the consideration could concern the existence of an
enclosure able to retain the products of the annihilation reaction. However, the assumption of
the existence of such an enclosure can be avoided by noticing that we deal here with the volume
(where the reaction occurs) versus the area (where the boundary is) problem. By making, if
necessary, a large number of copies of the system-antisystem pairs and letting them interact
in their common volume one can see that for the second law of thermodynamics to hold, the entropy
should not decrease in the volume (the volume entropy grows faster with the number of copies
than the contribution to the entropy from the boundary). It is the latter demand that is expressed
by Eqs.~(\ref{boundgeneral}) and (\ref{bound}).


Let us note that the bound in the form (\ref{bound}) may fail for containers where one dimension of size $d$ is much smaller
than the other two of size $D$. If $d$ is much smaller than the thermal wavelength $\lambda$ of the photons, while
$D\gg \lambda$, then we deal with effectively two-dimensional gas of photons where $S\propto (DE)^{2/3}$.
The latter value exceeds $(D^2 d)^{1/4} E^{3/4}$ following from (\ref{bound}) if $d$ is sufficiently small.
This and similar counterexamples to the bound can be avoided be rewriting the latter in its more fundamental
form
\begin{eqnarray}&&
S\leq {\hat S}_{rad}(E, V), \label{bound1}
\end{eqnarray}
where ${\hat S}_{rad}(E, V)$ is defined as the entropy of radiation that takes into account the given shape
of the container. In the form (\ref{bound1}) the bound holds also for special shapes of container such as
described above.

In the cases where gravity effects become important, the bound modifications are needed. In
particular, the PUW bound has a different form for small system volumes, $V\lesssim R_g^4 R^{-1}_{Comp}$, where
$R_g\equiv G E/c^4$ is the gravitational radius of the system \cite{Page}. Here the equilibrium
state of thermal radiation includes a black hole that exchanges radiation with the system via the Hawking
radiation process \cite{Hawking}. As a result, at these $V$ one should use for $S_{rad}(E, V)$ in
Eq.~(\ref{bound}) an expression that involves a sum of the entropies of the black hole and the remaining radiation, see \cite{Page}.

As with any entropy bound there is a question of how precisely the bound works and how the
situations where it could fail are avoided. An example of a challenge to entropy bounds is the zero
mode argument proposed in \cite{Unruh}. It is argued that a zero mode possibly present in a field
can accommodate arbitrarily large entropy at zero energy cost that would lead to a violation of an
entropy bound. This argument against a bound was refuted in \cite{SB} where it was shown that in
fact a zero mode contributes vanishing entropy to the system. A general discussion of how entropy
bounds work can be found in \cite{Bekenstein}.

The proposed gedanken experiment is informative about any system whatever the details of the
product of annihilation reaction are. However, it seems most interesting where one can say that
the product is radiation thus obtaining the PUW bound for the considered system. Then the physical
reason for the bound is the expected instability of the matter-antimatter
couple and the irreversibility of the resulting annihilation reaction. While radiation is
expected to result from annihilation rather generally, it seems impossible to exclude on general grounds the possibility of unusual situations with other reaction results such as bound matter-antimatter
states. In this case the bound in the form described by Eq.~(\ref{boundgeneral}) will still work and one would have to calculate the entropy of the bound state.

Thus the main drawback of the proposed consideration is the absence of a condition to decide the product of the annihilation reaction. It is clear however that the product is radiation for most known real systems. Thus the present work demonstrates the bound for a wide class of systems occurring in nature. Moreover, it shows for the first time what may go wrong if the bound does not hold. By establishing that the assumption that annihilation leads to radiation implies the bound we get a new point of view on the bound that might eventually lead to significant progress in its understanding. Below we indicate two
directions for future studies implied by the present work.

Our consideration shows that the bound can be expected to hold rather generally in theories
that incorporate antimatter. Then the CPT theorem (see e. g. \cite{Weinberg}) implies
that a local Lorentz invariant quantum field theory (below LLIQFT) obeys the bound (\ref{boundgeneral}).
Furthermore, in a general class of situations where the product of the annihilation reaction is
given by the radiation, a LLIQFT must also obey the PUW bound.
Using the Boltzmann formula for the entropy $S(E, V)=\ln g(E, V)\Delta E$, where $g(E, V)$ is
the density of states and $\Delta E$ is the energy window, we conclude that density of states
of a wide class of LLIQFT should grow with $E$ not faster than $\exp[S_{rad}(E, V)]$ (the factor in front of the
exponent gives a subleading dependence on $E$). This sheds new light on the calculations of \cite{Yurtsever,Aste},
aimed to demonstrate the validity of the holographic entropy bound for LLIQFT. The holographic bound
$S\leq \pi c^3 R^2/\hbar G$ was introduced by 't Hooft and Susskind \cite{Hooft,Susskind} and it can be obtained
from Eq.~(\ref{bound}) by using $R_g<R$. The above leads to the conjecture that under the assumptions of LLIQFT the PUW bound must hold. The study of this conjecture is a subject for further work.

Another direction that the present work suggests is the generalization of the PUW bound to include gravity. As mentioned in the introduction, the PUW bound does not hold in the situations with strong
gravity where one should rather use the Bekenstein bound (\ref{a1}). Is it possible to derive a more
general bound that would reproduce the two bounds in the limits of weak and strong gravity? Since the suggested gedanken experiment can be considered in situations where the self-gravity is not negligible,
it is not unlikely that such a generalized bound exists. The study of this possibility is an
important subject for further studies.

I am very grateful to J. D. Bekenstein for numerous helpful discussions and kind attitude. I also thank
very much S. Nussinov for a very fruitful discussion. A. Casher is acknowledged for a useful comment and M. Vucelja for help in preparing the
picture. This work was supported by DIP $0603215013$ and BSF $0603215611$ grants.

\end{document}